# Decoding Visual Responses based on Deep Neural Networks with Ear-EEG Signals

Young-Eun Lee[1], Minji Lee[1]
[1]Department of Brain and Cognitive Engineering, Korea University, Seoul, Republic of Korea
ye_lee@korea.ac.kr, minjilee@korea.ac.kr

*Abstract*—Recently, practical brain-computer interface is actively carried out, especially, in an ambulatory environment. However, the electroencephalography signals are distorted by movement artifacts and electromyography signals in ambulatory condition, which make hard to recognize human intention. In addition, as hardware issues are also challenging, ear-EEG has been developed for practical brain-computer interface and is widely used. However, ear-EEG still contains contaminated signals. In this paper, we proposed robust two-stream deep neural networks in walking conditions and analyzed the visual response EEG signals in the scalp and ear in terms of statistical analysis and brain-computer interface performance. We validated the signals with the visual response paradigm, steady-state visual evoked potential. The brain-computer interface performance deteriorated as 3~14% when walking fast at 1.6 m/s. When applying the proposed method, the accuracies increase 15% in cap-EEG and 7% in ear-EEG. The proposed method shows robust to the ambulatory condition in session dependent and session-to-session experiments.

## I. INTRODUCTION

Brain-computer interfaces (BCIs) in ambulatory conditions is one of the most important factors in real life. While many have studied BCIs for detecting human cognitive condition or intention based on electroencephalography (EEG), limitations used in real life are exposed to the ambulatory environment [l, 2]. This is because the EEG signals are distorted much in ambulatory conditions [3, 4]. Movement artifacts are caused by electromyography generated by muscle activity when walking, or by skin and cable movements. Since these artifacts are much larger in amplitude than the brain signal that contains the user's intent, it is difficult to catch properly the meaning of intention. Therefore, decoding human intention in the ambulatory environment is tried these days [4-7] using movement artifact removal methods [8, 9] and deep neural networks [7, 10-17] to robust the artifacts and increase the performance.

The development of simple hardware to measure EEG signals has become a big issue. Among them, the ear-EEG has recently been studied extensively by many researchers to improve user convenience and validated by analyzing the signal quality and executing various BCI paradigms. In addition, the conventional hardware to measure EEG is annoying users in terms of high cost and difficulties to setup. Setup of EEG cap uses a conductive gel on the hairs, which needs to wash after measuring, and stands out due to wearing an uncomfortable cap. In order to reduce the tiresome, uncomplicated devices were designed, such as Emotive EPOC and ear-EEG. In Kidmose et al. [18], the researchers analyzed scalp and ear-EEG signals with steady-state and transient event-related potential (ERP) paradigms. Moreover, Debener et al. [19, 20] designed cEEGrid which placed electrodes around the ear. The cEEGrid could preserve the ERP signals and have similar performance to cap-EEG. Whereas, there is a limitation that the performance is lower than conventional scalp-EEG. Therefore, several studies tried to increase the performance of ear-EEG for visual or auditory responses [21].

BCI paradigms are mainly developed to motor imagery [22-27], ERP [28-32], and steady-state visual evoked potential (SSVEP) [6, 7, 32-34]. ERP and SSVEP are visual responses and are widely used to recognize human intention because their patterns in EEG signals are relatively huge and they showed reliable performance when it comes to accuracy and response time with only a few EEG channels comparing to other BCI paradigms. SSVEPs and ERPs including P300 are primary used visual response paradigms. SSVEPs are the natural visual responses evoked in the occipital cortex to periodic visual stimuli at specific frequencies. They typically operate several targets at specific frequencies between 1 and 100 Hz and can be distinguished by their characteristic composition of harmonic frequencies [33].

There are many machine learning methods to recognize human intention through visual responses. For SSVEP, many use classifiers named canonical correlation analysis (CCA) [35, 36], multivariate statistical analysis for inferring information from cross-covariance matrices of the relationships among the variables. And these days, using machine learning methods and deep learning methods such as convolutional neural networks (CNN) could increase the accuracy of huge degrees in several papers [37]. In Kwak et al. [7], CNN are used to classify the visual response from cap-EEG, having 94.03% accuracy. This paper experimented in ambulatory conditions as well, riding exoskeleton. Another study [10] also used the features based on fast Fourier transform and CNN classifiers to recognize the human intention from SSVEP. The Castermans et al. [4] classified ERP intention in the ambulatory environment, up to 1.25 m/s using linear discriminant analysis (LDA) classifier.

This work was supported by Institute for Information & Communications Technology Promotion (IITP) grant funded by the Korea government (No. 2017-0-00451, Development of BCI based Brain and Cognitive Computing Technology for Recognizing User's Intentions using Deep Learning).

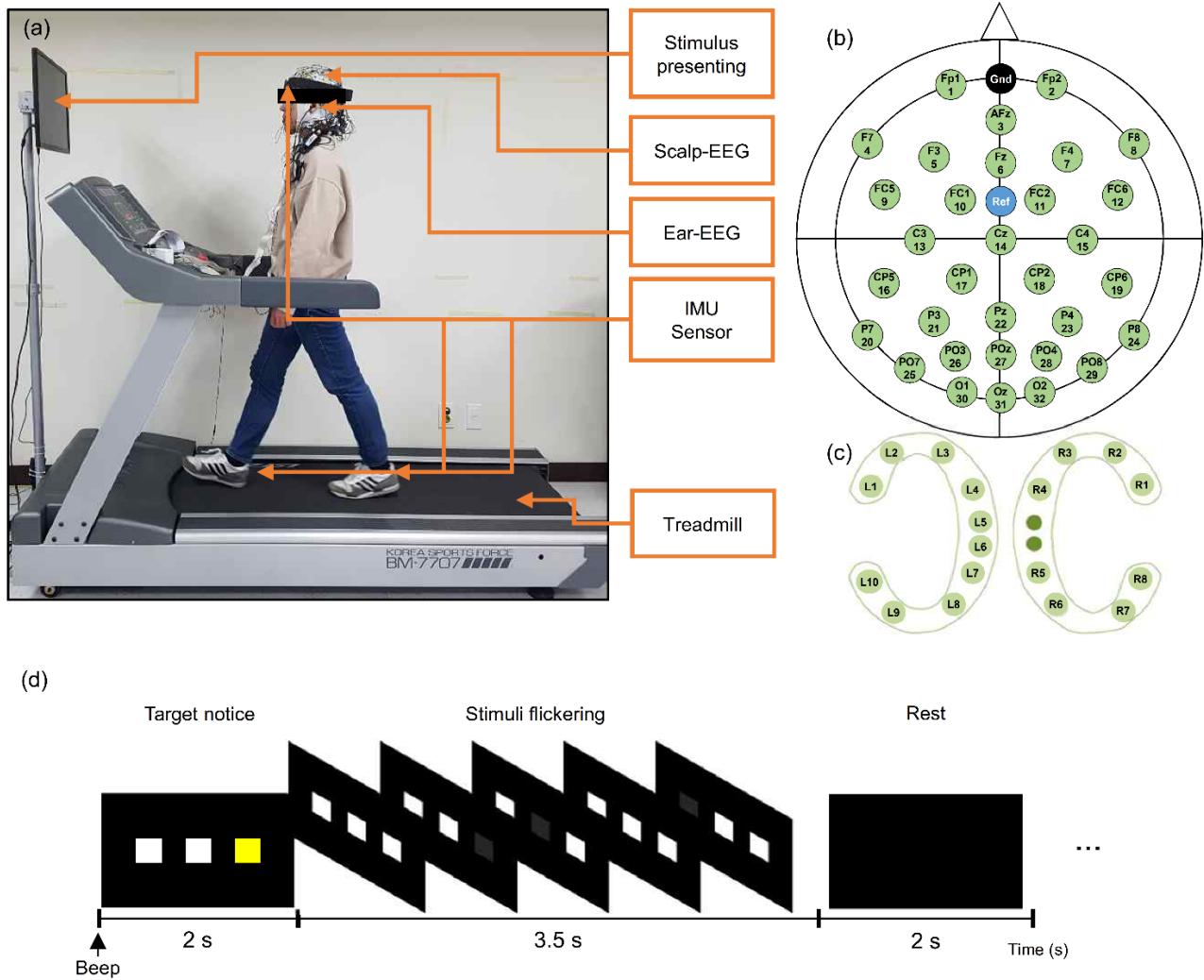

Figure 1. Experimental setup. (a) Experimental design. Subjects walked on a treadmill and focus on the presented stimulus on a screen. EEG signals from scalp and ear were measured and signals from IMU sensors were also collected at the same time. (b) Scalp-EEG channel placement. 32-channels focusing on occipital regions were collected. (c) Ear-EEG channel placement. 18 channels, 10 channels from left and 8 channels from right were measured. (d) Presented stimulus. Three different stimulus were flickering in 5.45, 8.75, and 12 Hz.

In this paper, we decoded the visual responses from cap-EEG and ear-EEG in the ambulatory environment. For practical BCIs, simple hardware and high accurate classifier of human intention is necessary. Therefore, we investigated deep learning methods to increase human intention recognition and used ear-EEG for practical BCI in the real-world.

## II. MATERIALS AND METHODS

### A. Participants

We included thirteen healthy participants (3 females, age 24.5 ±2.5 years) with normal or corrected-to-normal vision and no difficulties to walk at Korea University in Seoul, Korea. None of the participants had a history of neurological, psychiatric, or any other pertinent disease that otherwise might have affected the experimental results. This study was reviewed and approved by the Korea University Institutional Review Board (KUIRB-2019-0194-01).

### B. Experimental Paradigm

The subjects were stood on the treadmill at 80 (±5) cm in front of a monitor and walked at two different speeds (0.8 and 1.6 m/s). We experimented with three target frequencies SSVEP paradigm. Participants were maintained from the screen at 80 (±10) cm in front of a 60 Hz LCD monitor (Samsung, SyncMaster 2494HM, refresh rate: 60 Hz; resolution: 1920 × 1080). The white-colored SSVEP stimuli were designed to flicker at 5.45, 6.67, and 8.57 Hz, which were calculated by dividing the monitor refresh rate by an integer (60/11, 60/7, and 60/5). The size of the stimuli was 6 cm × 6 cm. Each stimulus was presented for 3.5 s with an inter-stimulus interval of 4 s. The visual stimuli were generated using the Psychophysics Toolbox in Matlab. We followed the SSVEP paradigm of a previous study [38].

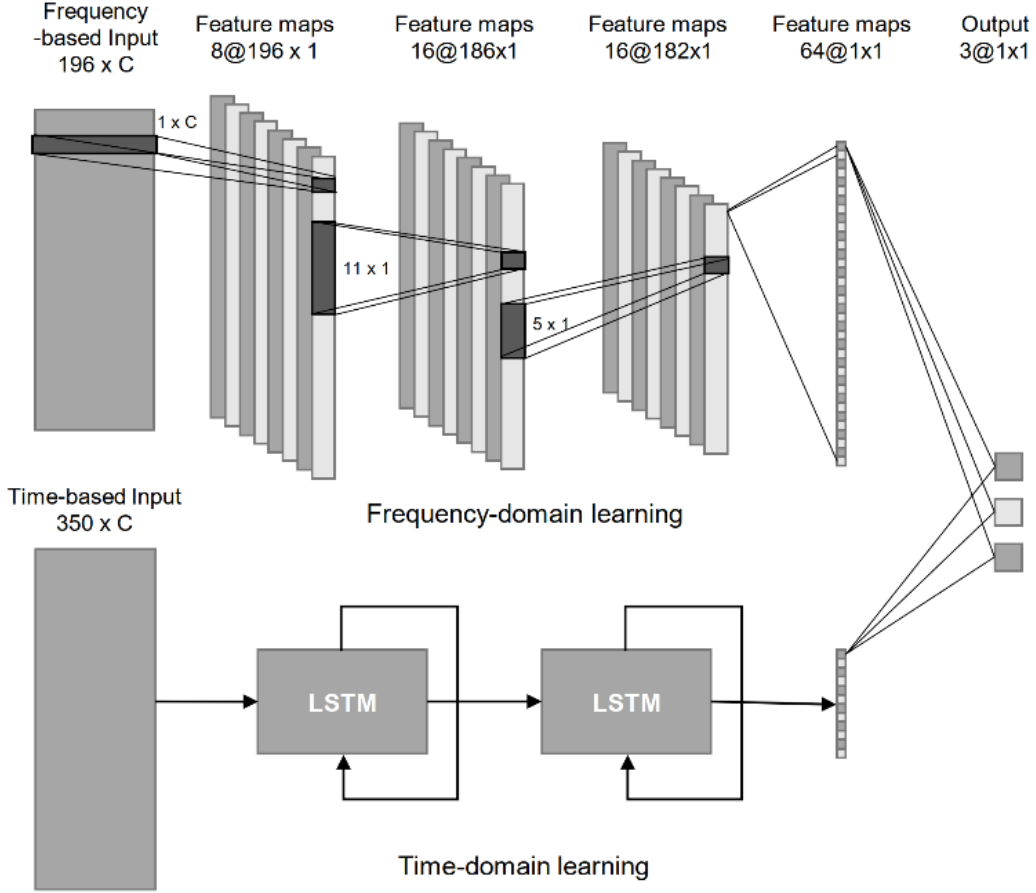

Figure 2. Deep learning networks structure. Two types of input, frequency-based input and time-based input are used. Frequency-domain learning has four hidden layers and time-domain learning has three hidden layers. C indicated the number of channels, according to scalp and ear-EEG.

*C. Data Acquisition and Preprocessing*

Figure 1 shows the experimental setup about measurement, experimental tools, and channel placement. We recorded 32-channel of cap-EEG, 18-channels of ear-EEG, and 9-channels Inertial Measurement Unit (IMU) sensors. We used a wireless interface (MOVE system, Brain Product GmbH) and Ag/AgCl electrodes to acquire EEG signals from the scalp and Smarting System (mBrainTrain LLC) and cEEGrid electrodes to acquire EEG signals from ear. Three wearable IMU sensors have recorded the movement on the head, left and right ankles. The cap electrodes were placed according to the 10-20 international system at locations: Fp1, Fp2, AFz, F7, F3, Fz, F4, F8, FC5, FC1, FC2, FC6, C3, Cz, C4, CP5, CP1, CP2, CP6, P7, P3, Pz, P4, P8, PO7, PO3, POz, PO4, PO8, O1, Oz, and O2. Ear-EEG electrodes were cEEGrid, having 10 channels on left side (L1 to L10), 8 channels on right side (R1 to R8) and GND and REF in the middle of right side. The impedances were maintained below 10 kΩ for both scalp and ear-EEG. We set the sampling rate as 500 Hz for cap and ear-EEG and 128 Hz for IMU sensors.

All BCI experiments were developed based on the OpenBMI [39], BBCI [40] and Psychophysics toolboxes [41]. We performed down sampling or resampling to 100 Hz for all measurement and high-pass filter using finite impulse response filter passing above 3 Hz.

*D. Proposed Deep Neural Networks*

We utilized two different features, frequency-domain features, and time-domain features, to train deep neural networks algorithm described in figure 2. Two-stream deep neural networks are used, having CNNs for frequency-domain feature training and long-short term memory (LSTM) for time-domain feature training. For acquiring frequency-based input, FFT is used for each channel and is denoted by

$$X_f = \sum_{t=0}^{T-1} e^{-2\pi i f\left(\frac{t}{T}\right)} x_t \quad (1)$$

where $x_t$ is time-domain input, $T$ is time, $X_f$ is frequency-domain output. The neural networks for frequency domain are three CNN-hidden layers and a fully-connected hidden layers. In the first layer, eight kernels, channel-wise convolution, having a size of 1 by the number of channels are used and the feature maps have a shape of time by 1. The feature maps are calculated by

$$x_k = f(\sigma_k(p)) \quad (2)$$

where $\sigma_k$ is convolutional function, $p$ is a position of input, $f(\cdot)$ function is activation function, rectified linear unit (ReLU) is used for the activation function and is denoted by

$$f(z) = \max(0, z) \quad (3)$$

TABLE I. TABLE 1 CLASSIFICATION ACCURACY OF EAR-EEG FOR ALL SUBJECTS USING DIFFERENT METHODS IN DIFFERENT WALKING ENVIRONMENT.

| Speed | Methods | S1 | S2 | S3 | S4 | S5 | S6 | S7 | S8 | S9 | S10 | S11 | S12 | S13 | Average | SD |
|---|---|---|---|---|---|---|---|---|---|---|---|---|---|---|---|---|
| **Standing** | CCA[a] | 0.40 | 0.62 | 0.42 | 0.73 | 0.70 | 0.45 | 0.47 | 0.42 | 0.57 | 0.47 | 0.37 | 0.57 | 0.47 | 0.51 | 0.12 |
| | LDA[b] | 0.38 | 0.46 | 0.41 | 0.41 | 0.30 | 0.35 | 0.27 | 0.41 | 0.27 | 0.30 | 0.46 | 0.35 | 0.38 | 0.36 | 0.06 |
| | Proposed[c] | 0.42 | 0.38 | 0.38 | 0.63 | 0.50 | 0.54 | 0.71 | 0.42 | 0.38 | 0.38 | 0.54 | 0.58 | 0.54 | 0.49 | 0.11 |
| **0.8m/s** | CCA | 0.37 | 0.38 | 0.53 | 0.48 | 0.53 | 0.38 | 0.42 | 0.38 | 0.43 | 0.38 | 0.32 | 0.47 | 0.38 | 0.42 | 0.07 |
| | LDA | 0.32 | 0.35 | 0.46 | 0.38 | 0.41 | 0.24 | 0.43 | 0.43 | 0.54 | 0.35 | 0.35 | 0.32 | 0.32 | 0.38 | 0.08 |
| | Proposed | 0.42 | 0.38 | 0.42 | 0.79 | 0.54 | 0.46 | 0.42 | 0.50 | 0.67 | 0.38 | 0.50 | 0.46 | 0.46 | 0.49 | 0.12 |
| **1.6m/s** | CCA | 0.35 | 0.52 | 0.43 | 0.37 | 0.50 | 0.42 | 0.33 | 0.45 | 0.38 | 0.38 | 0.47 | 0.37 | 0.37 | 0.41 | 0.06 |
| | LDA | 0.22 | 0.49 | 0.43 | 0.41 | 0.41 | 0.43 | 0.46 | 0.57 | 0.43 | 0.38 | 0.62 | 0.32 | 0.41 | 0.43 | 0.10 |
| | Proposed | 0.50 | 0.33 | 0.46 | 0.71 | 0.46 | 0.50 | 0.46 | 0.50 | 0.58 | 0.46 | 0.50 | 0.38 | 0.42 | 0.48 | 0.09 |

a. CCA = canonical correlation analysis
b. LDA = linear discriminant analysis
c. Proposed method = convolutional neural networks + long-short term memory

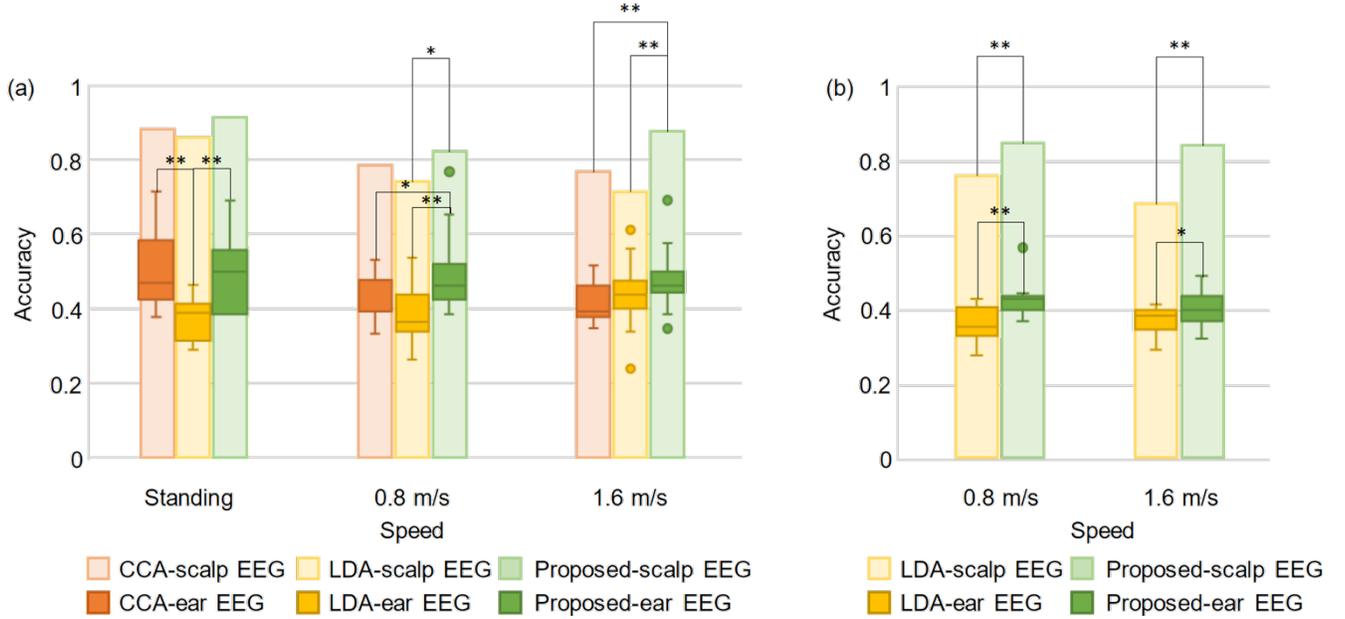

Figure 3. Performance of each method in different speeds. Left graph (a) shows the accuracy of each sessions in different speeds. Bar graphs represent the accuracies of EEG from scalp and the box graphs represent the average and standard deviation of the accuracies of EEG from ear. Right graph (b) shows the accuracy of session-to-session learning in different speeds. Training data was standing data for each subjects and test data was walking data in each speed. One asterisk indicates the 5% significance level between accuracies of two methods and two asterisks indicate the 1% significance level.

The convolutional function $\sigma_k$ can be represented by

$$\sigma_k(p) = b_k + \sum_{i=1}^{i=K_x} \sum_{j=1}^{j=K_y} x^{i,j} \cdot w_k^{i,j} \quad (4)$$

where $b_k$ is bias of kernel $k$, $K_x$ and $K_y$ is kernel size, $x$ is input matrix, and $w_k$ is weight of kernel $k$.

Time-domain features were gathered by LSTM having three hidden layers. After acquiring feature maps, we combined two feature maps, time-domain and frequency-domain features.

Cross-entropy loss, logarithm multiplying classes, are used for the loss function of the algorithm, which is denoted by

$$loss = -t\log(f(s)) - (1-t)\log(1 - f(s)) \quad (5)$$

where $t$ is a class in [0,1], $s$ is the ground truth, and $f(s)$ is prediction function. The learning rate was 0.01 and weights were initialized with a normal distribution. The number of epochs was 50 and batch size was 32.

III. RESULTS

For evaluating our proposed methods, we compared with conventional methods, CCA and LDA. And we analyzed data for session dependent and session-to-session from standing to walking sessions. As CCA is a statistical analysis method that doesn't need training data, we analyzed session-to-session data using only LDA and proposed method. The results were statistically analyzed using the statistical method of t-test. Table I shows the results of analysis for SSVEP from scalp and ear-

EEG and indicates all subjects' performance for each method in each different walking speed.

## A. Session Dependent Classification

Figure 3 (a) indicates the accuracy of each method from scalp and ear-EEG for independent sessions. The bar graphs represent the accuracy of data from scalp-EEG. In session independent analysis, there are a few decreases as increasing the speed of walking for scalp-EEG data. When we used CCA or LDA for classification, the classification performance of 1.6 m/s data decreased from standing data as 11% and 14%, respectively. The proposed method for 1.6 m/s, however, decreased performance as only 4% comparing standing data ($p_{CCA-CNN}$ = 0.045, $p_{LDA-CNN}$ = 0.023). The box graphs represent the average and standard deviation of the accuracy of data from ear-EEG. The performance for ear-EEG in 1.6 m/s using CCA decreased as 10%. On the other hand, the proposed method decreased as only 1%. Moreover, in speed 1.6 m/s, the proposed method had the highest performance for both scalp ($p$ = 0.004) and ear-EEG.

## B. Session-to-session Classification

Figure 3 (b) indicates the accuracy of each method from scalp and ear-EEG for session-to-session. We used EEG data in standing condition as training data and walking in 0.8 m/s and 1.6 m/s as test data. The session-to-session accuracy is lower than session-dependent, especially in ear-EEG. The accuracies using proposed method for scalp-EEG in 1.6 m/s using LDA are much higher than LDA having 66% and 81%, respectively ($p$ < 0.001). The accuracies using proposed method for ear-EEG in 1.6m/s are also higher than LDA having 35% and 39%, respectively ($p$ = 0.047).

## IV. DISCUSSION AND CONCLUSION

In this study, we proposed a robust method in walking condition decoding visual response in scalp and ear-EEG using deep neural networks. Practical BCIs require a robust system in an ambulatory environment and simple hardware usable in the real-world. We decoded visual responses such as SSVEP from scalp and ear-EEG in different walking conditions. The proposed method is two-stream deep learning architecture using frequency-domain features and time-domain features. We show that the proposed method is robust in the ambulatory environment comparing with other methods.

Kwak et al. [7] was reported that the feature map from neural networks had strong channel-wise features and frequency-wise features so that the networks could distinguish the intention of humans in noisy conditions. Therefore, the results for recognizing human intention in an ambulatory environment without using artifact removal methods had the reasonable performance in session-dependent data. However, for session-to-session data, the performances were lower that could not fit network because of the artifacts' variance.

In conclusion, we showed that deep neural networks could show reasonable performance for each session even in the ambulatory environment. However, it was difficult to increase the performance fitting from training data in standing condition to test data in walking condition due to huge artifacts features. In the future, the study removing noisy signals but remaining essential components was necessary to recognize the human intention for a different session in the ambulatory environment.

## REFERENCES

[1] T. P. Luu, S. Nakagome, Y. He, and J. L. Contreras-Vidal, "Real-time EEG-based brain-computer interface to a virtual avatar enhances corticalinvolvement in human treadmill walking," *Sci. Rep.*, vol. 7, pp. 8895, Aug. 2017.

[2] T. P. Luu, Y. He, S. Nakagame, J. Gorges, K. Nathan, and J. L. Contreras-Vidal, "Unscented kalman filter for neural decoding of human treadmill walking from non-invasive electroencephalography," *Int. Conf. Proc. IEEE Eng. Med. Biol. Soc. (EMBC)*, pp. 1548–1551, Aug. 2016.

[3] B. R. Malcolm, J. J. Foxe, J. S. Butler, W. B. Mowrey, S. Molholm, and P. De Sanctis, "Long-term test-retest reliability of event-related potential (ERP) recordings during treadmill walking using the mobile brain/ bodyimaging (MoBI) approach," *Brain. Res.*, vol. 1716, pp. 62-69, Aug. 2019.

[4] T. Castermans, M. Duvinage, M. Petieau, T. Hoellinger, C. De Saedeleer, K. Seetharaman, A. Bengoetxea, G. Cheron, and T. Dutoit, "Optimizing the performances of a p300-based brain–computer interface in ambulatory conditions," *IEEE. J. Emerg. Sel. Top. Circuits. Syst.*, vol. 1, pp. 566–577, Dec. 2011.

[5] K. Gramann, J. T. Gwin, N. Bigdely-Shamlo, D. P. Ferris, and S. Makeig, "Visual evoked responses during standing and walking," *Front. Hum. Neurosci.*, vol. 4, pp. 1-12, Oct. 2010.

[6] N.-S. Kwak, K.-R. Müller, and S.-W. Lee, "A lower limb exoskeleton control system based on steady state visual evoked potentials," *J. Neural. Eng.*, vol. 12, pp. 056009, Oct. 2015.

[7] N.-S. Kwak, K.-R. Müller, and S.-W. Lee, "A convolutional neural network for steady state visual evoked potential classification under ambulatory environment," *PLoS One*, vol. 12, pp. e0172578, Feb. 2017.

[8] T. C. Bulea, J. Kim, D. L. Damiano, C. J. Stanley, and H.-S. Park, "Prefrontal, posterior parietal and sensorimotor network activity underlying speed control during walking," *Front. Hum. Neurosci.*, vol. 9, pp. 1-13, May 2015.

[9] A. D. Nordin, W. D. Hairston, and D. P. Ferris, "Dual-electrode motion artifact cancellation for mobile electroencephalography," *J. Neural. Eng.*, vol. 15, pp. 056024, Aug. 2018.

[10] T.-H. Nguyen and W.-Y. Chung, "A single-channel SSVEP-based BCI speller using deep learning," *IEEE Access*, vol. 7, pp. 1752-1763, Dec. 2018.

[11] H. H. Bülthoff, S.-W. Lee, T. A. Poggio, and C. Wallraven, Biologically Motivated Computer Vision. *NY: Springer-Verlag*, 2003.

[12] M. Kim, G. Wu, Q. Wang, S.-W. Lee, and D. Shen, "Improved image registration by sparse patch-based deformation estimation," *Neuroimage*, vol. 105, pp. 257-268, Jan. 2015.

[13] X. Ding and S.-W. Lee, "Changes of functional and effective connectivity in smoking replenishment on deprived heavy smokers: a resting-state FMRI study," *PLoS One*, vol. 8, pp. e59331, Mar. 2013.

[14] I.-H. Kim, J.-W. Kim, S. Haufe, and S.-W. Lee, "Detection of braking intention in diverse situations during simulated driving based on EEG feature combination," *J. Neural Eng.*, vol. 12, pp. 016001, Nov. 2014.

[15] X. Chen, H. Zhang, L. Zhang, C. Shen, S.-W. Lee, and D. Shen, "Extraction of dynamic functional connectivity from brain grey matter and white matter for MCI classification," *Hum. Brain Mapp.*, vol. 38, pp. 5019-5034, Jun. 2017.

[16] M. Lee, R. D. Sanders, S.-K. Yeom, D.-O. Won, K.-S. Seo, H.-J. Kim, G. Tononi, and S.-W. Lee, "Network properties in transitions of consciousness during propofol-induced sedation," *Sci. Rep.*, vol. 7, pp. 16791, Dec. 2017.

[17] O.-Y. Kwon, M.-H. Lee, C. Guan, and S.-W. Lee, "Subject-independent brain-computer interfaces based on deep convolutional neural networks," *IEEE Trans. Neural Netw. Learn. Syst.*, E-pub, pp. 1-15, Nov. 2019.

[18] P. Kidmose, D. Looney, M. Ungstrup, M. L. Rank, and D. P. Mandic, "A study of evoked potentials from ear-EEG," *IEEE. Trans. Biomed. Eng.*, vol. 60, pp. 2824-2830, May. 2013.

[19] S. Debener, R. Emkes, M. De Vos, and M. Bleichner, "Unobtrusive ambulatory EEG using a smartphone and flexible printed electrodes around the ear," *Sci. Rep.*, vol. 5, pp. 1-11, Nov. 2015.

[20] M. G. Bleichner and S. Debener, "Concealed, unobtrusive ear-centeredeeg acquisition: ceegrids for transparent EEG," *Front. Hum. Neurosci.*, vol. 11, pp. 163, Apr. 2017.

[21] N.-S. Kwak and S.-W. Lee, "Error correction regression framework for enhancing the decoding accuracies of ear-EEG brain-computer interfaces," *IEEE Trans. Cybern.*, pp. 1-14, Jul. 2019.

[22] H.-I. Suk and S.-W. Lee, "Subject and class specific frequency bands selection for multiclass motor imagery classification," *Int. J. Imaging Syst. Technol.*, vol. 21, pp. 123-130, May. 2011.

[23] M. Lee, C.-H. Park, C.-H. Im, J.-H. Kim, G.-H. Kwon, L. Kim, W.-H. Chang, and Y.-H. Kim, "Motor imagery learning across a sequence of trials in stroke patients," *Restor. Neuro. Neurosci.*, vol. 34, pp. 635-645, Aug. 2016.

[24] R. T. Schirrmeister, J. T. Springenberg, L. D. J. Fiederer, M. Glasstetter, K. Eggensperger, M. Tangermann, F. Hutter, W. Burgard, and T. Ball, "Deep learning with convolutional neural networks for EEG decoding and visualization," *Hum. Brain Mapp.*, vol. 38, pp. 5391-5420, Nov. 2017.

[25] J.-H. Kim, F. Bießmann, and S.-W. Lee, "Decoding three-dimensional trajectory of executed and imagined arm movements from electroencephalogram signals," *IEEE Trans. Neural Syst. Rehabil. Eng.*, vol. 23, pp. 867-876, Dec. 2014.

[26] M.-H. Lee, S. Fazli, J. Mehnert, and S.-W. Lee, "Subject-dependent classification for robust idle state detection using multi-modal neuroimaging and data-fusion techniques in BCI," *Pattern. Recognit.*, vol. 48, pp. 2725-2737, Aug. 2015.

[27] T.-E. Kam, H.-I. Suk, and S.-W. Lee, "Non-homogeneous spatial filter optimization for ElectroEncephaloGram(EEG)-based motor imagery classification," *Neurocomputing*, vol. 108, pp. 58-68, May. 2013.

[28] S.-K. Yeom, S. Fazli, K.-R. Müller, and S.-W. Lee, "An efficient ERP-based brain-computer interface using random set presentation and face familiarity," *PLoS One*, vol. 9, pp. e111157, Nov. 2014.

[29] Y. Chen, A. D. Atnafu, I. Schlattner, W. T. Weldtsadik, M.-C. Roh, H.-J. Kim, S.-W. Lee, B. Blankertz, and S. Fazli, "A high-security EEG-based login system with RSVP stimdetectinguli and dry electrodes," *IEEE Trans. Inf. Forensic Secur.*, vol. 11, pp. 2635-2647, Jun. 2016.

[30] D.-O. Won, H.-J. Hwang, D.-M. Kim, K.-R. Müller, and S.-W. Lee, "Motion-based rapid serial visual presentation for gaze-independent brain-computer interfaces," *IEEE Trans. Neural Syst. Rehabil. Eng.*, vol. 26, pp. 334-343, Feb. 2018.

[31] M.-H. Lee, J. Williamson, D.-O. Won, S. Fazli, and S.-W. Lee, "A high performance spelling system based on EEG-EOG signals with visual feedback," *IEEE Trans. Neural Syst. Rehabil. Eng.*, vol. 26, pp. 1443-1459, Jul. 2018.

[32] M.-H. Lee, J. Williamson, Y.-E. Lee, and S.-W. Lee, "Mental fatigue in central-field and peripheral-field steady-state visually evoked potential and its effects on event-related potential responses," *Neuroreport*, vol. 29, pp. 1301–1308, Oct. 2018.

[33] G.-R. Müller-Putz, R. Scherer, C. Neuper, and G. Pfurtscheller, "Steady-state somatosensory evoked potentials: suitable brain signals for brain-computer interfaces?," *IEEE. Trans. Neural. Syst. Rehabil. Eng.*, vol. 14, pp. 30-37, Mar. 2006.

[34] D.-O. Won, H.-J. Hwang, S. Dähne, K.-R. Müller, and S.-W. Lee, "Effect of higher frequency on the classification of steady-state visual evoked potentials," *J. Neural. Eng.*, vol. 13, pp. 016014, Feb. 2016.

[35] Z. Lin, C. Zhang, W. Wu, and X. Gao, "Frequency recognition based on canonical correlation analysis for SSVEP-based BCIs," *IEEE. Trans. Biomed. Eng.*, vol. 53, pp. 1172-1176, May. 2007.

[36] X. Zhu, H.-I. Suk, S.-W. Lee, and D. Shen, "Canonical feature selection for joint regression and multi-class identification in Alzheimer's disease diagnosis," *Brain Imaging Behav.*, vol. 10, pp. 818-828, Sep. 2016.

[37] M. Lee, S.-K. Yeom, B. Baird, O. Gosseries, J. O. Nieminen, G. Tononi, and S.-W. Lee, "Spatio-temporal analysis of EEG signal during consciousness using convolutional neural network," *Int. Conf. Proc. IEEE Brain-Computer Interface (BCI)*, pp. 1-3, Jan 2018.

[38] M.-H. Lee, O.-Y. Kwon, Y.-J. Kim, H.-K. Kim, Y.-E. Lee, J. Williamson, S. Fazli, and S.-W. Lee, "EEG dataset and openbmi toolbox for three bci paradigms: an investigation into bci illiteracy," *GigaScience*, vol. 8, pp. giz002, May. 2019.

[39] M.-H. Lee, S. Fazli, K.-T. Kim, and S.-W. Lee, "Development of an open source platform for brain-machine interface: OpenBMI," *Int. Conf. Proc. IEEE Brain-Computer Interface (BCI)*, pp. 1-2, Feb. 2016.

[40] R. Krepki, B. Blankertz, G. Curio, and K.-R. Müller, "The berlin brain-computer interface (BBCI) – towards a new communication channel foronline control in gaming applications," *Multimed. Tools Appl.*, vol. 33, pp. 73–90, Apr. 2007.

[41] M. Kleiner, D. Brainard, and D. Pelli, "What's new in psychtoolbox-3?," *Perception*, vol. 36, pp. 1-16, Aug. 2007.